\documentclass[aps,pre,showpacs,floatfix]{revtex4-1}
\usepackage{amsmath}
\usepackage{graphicx}
\usepackage{color}
\usepackage{amssymb}

\def\beq{\begin{eqnarray}}
\def\eeq{\end{eqnarray}}

\begin{document}

\title{Comment to "Thomson rings in a disk"}
\author{Paolo Amore}
\affiliation{Facultad de Ciencias, CUICBAS, Universidad de Colima, \\
Bernal D\'{\i}az del Castillo 340, Colima, Colima, M\'exico}

\begin{abstract}
We have found that the minimum energy configuration of $N=395$ charges confined in a disk and interacting via the Coulomb potential, 
reported by Cerkaski et al. in Ref.~\cite{Cerkaski15} is not a global minimum of the total electrostatic energy.
We have identified a large number of configurations with lower energy, where defects are present close to the center of
the disk; thus, the formation of a hexagonal core and valence circular rings for the centered configurations, predicted by the
model of Ref.~\cite{Cerkaski15}, is not supported by numerical evidence and the configurations obtained with this model 
cannot be used as a guide for the numerical calculations, as claimed by the authors. 
\end{abstract}

\maketitle



In a recent paper, ref.~\cite{Cerkaski15}, Cerkaski et al. studied the problem of a finite number of equal charges, interacting
via the Coulomb potential and confined inside a disk. This problem has been previously studied by several authors in a series 
of papers, refs.~\cite{Berezin85,Wille85,Nurmela98,Oymak00,Oymak01,Worley06,Moore07}, and it can be regarded as a 
generalization of the well--known Thomson problem \cite{Thomson} (finding the configurations of minimum energy
of $N$ equal charges on the surface of a sphere). Despite the apparent simplicity, both problems provide a serious
computational challenge, of increasing difficulty with $N$: in particular, the number of local minima 
of the total electrostatic energy grows very fast with $N$ (for the case of the Thomson problem see for example the discussion in
Ref.~\cite{Calef15}). As a result, the identification of the global 
minimum of a system of $N$ charges typically requires extensive numerical calculations: in the absence of a formal criterium 
to establish whether a given configuration of equilibrium is a global minimum,  one has to repeat the numerical calculations 
several times, keeping $N$ fixed, and regard the  configuration with lowest energy among those obtained as a probable 
candidate for a global minimum. 

For the case of the disk, Erkoc and Oymak \cite{Oymak00, Oymak01} have observed the tendency, for systems with modest number
of charges ($N \leq 109$), to accomodate the charges on concentric rings, empirically deducing the rules for the distribution 
of charges on the disk (incidentally, most of the energies reported by these authors in Tables \cite{Oymak01} do not 
correspond to global minima).  The analysis performed by Cerkaski et al. in Ref.~\cite{Cerkaski15} is a refinement of the work of 
Erkoc and Oymak \cite{Oymak00, Oymak01}, and it relies on the hypothesis that charges arrange on concentric rings, for 
configurations of minimum energy. In this way, the original problem is reduced to the much simpler problem of calculating 
the electrostatic energy due to $p$ rings, each carrying an appropriate number of charges, uniformely distributed over the ring;
the equilibrium configuration in this case is obtained by solving a system of two equations (their eqs. (16) and (17)).

To test their model the authors have performed numerical (molecular dynamics, MD for short) calculations for systems up 
to $N=400$ charges. Based on this analysis they conclude that their approach allows one "to determine with high accuracy 
the equilibrium configurations of a few hundred charged particles". In particular,  for $N \gtrsim 200$ their approach 
"predicts the formation of the hexagonal core and valence circular rings for the centered configurations", with "an
 increasing sequence of rings, starting from the center, matching the regular hexagonal pattern". 
 For the case of $N=395$, discussed at length in Ref.~\cite{Cerkaski15} the authors observe the formation
 of an hexagonal structure with rings $\left\{1,6,12,18,24\right\}$. Fig. 2b of Ref.~\cite{Cerkaski15}, that reports a comparison between
 the model and the numerical MD calculations for $N=395$ charges, displays an excellent agreement between the two, with only a 
 small mismatch just outside the hexagonal structure (the green region in the figure is used to highlight the hexagonal
 structure). The energy reported by the authors for this configuration, which is expected to be a global minimum of the 
 total energy, is $\mathcal{E}_{MD} = 110665.1$, compared to the energy $\mathcal{E}_{avg} = 110667.6$, obtained with their model,
 with an error of just $2 \times 10^{-3} \%$.

\begin{figure}
\begin{center}
\bigskip\bigskip\bigskip
\includegraphics[width=7cm]{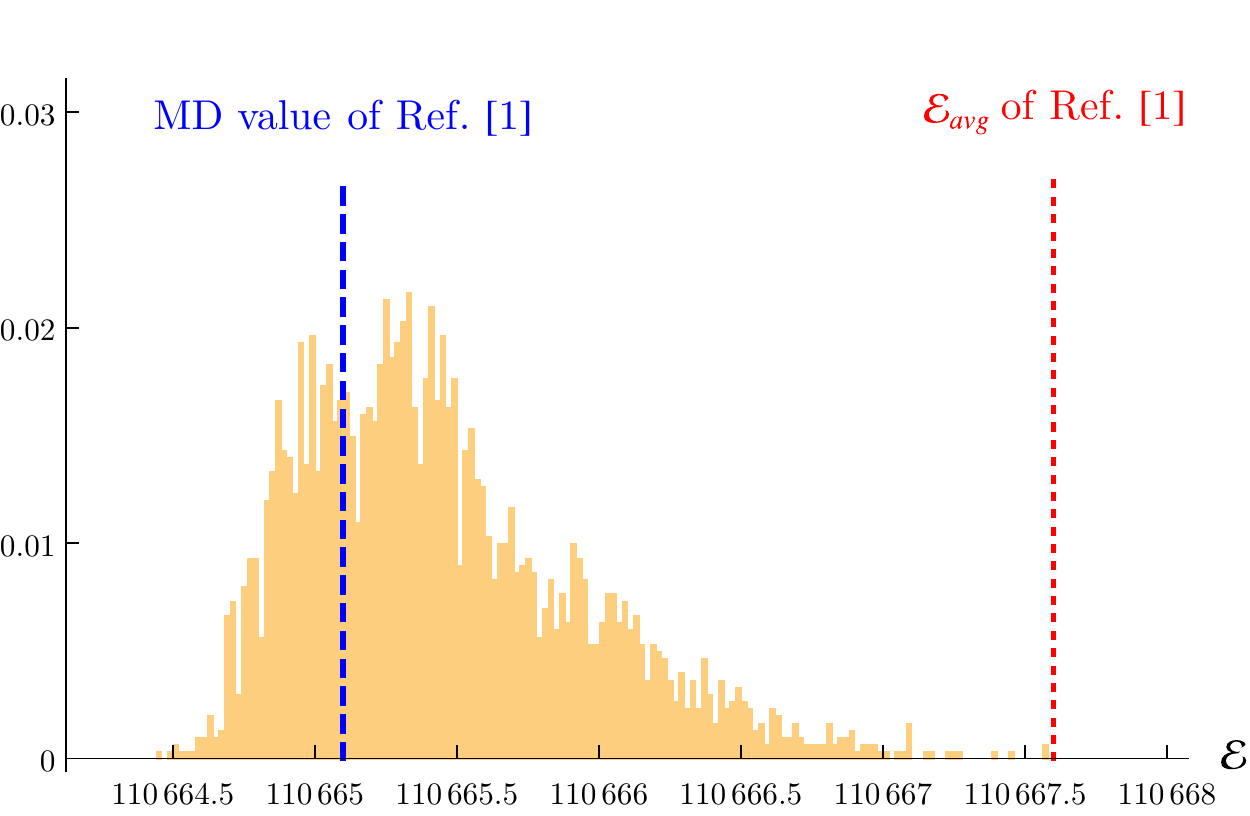} 
\caption{(color online) Histogram for the energies of the configurations with $395$ charges with $N_p=147$.}
\label{Fig_1}
\end{center}
\end{figure}

With the purpose of veryfying the results of Ref.~\cite{Cerkaski15} we have carried out extensive numerical calculations,
in particular for the case of $395$ charges. The approach that we have implemented allows one to generate configurations
with the desired number of charges on the border: in this way we have verified that the lower energies
occurr when $N_p=147$ charges are disposed on the border of the disk, in agreement with Ref.~\cite{Cerkaski15}.
We have thus generated $3001$ configurations with $N_p=147$, starting from initial configurations where the internal
charges are randomly distributed, and we have found that $824$ of them have energy lower that the value reported in 
Ref.~\cite{Cerkaski15}, $\mathcal{E} < \mathcal{E}_{MD} = 110665.1$. The lowest energy among those that we have 
calculated (possibly a global minimum) is $\mathcal{E}_{\rm MIN} = 110664.44$. 
The histogram in Fig.~\ref{Fig_1} illustrates these points.
Interestingly, we have also found that  even the configuration with largest energy has an energy slightly 
lower that the value predicted by the model of Ref.~\cite{Cerkaski15}, $\mathcal{E} = 110667.576 < \mathcal{E}_{avg}$ (see 
Fig.~\ref{Fig_1}).

\begin{figure}
\begin{center}
\bigskip\bigskip\bigskip
\includegraphics[width=8cm]{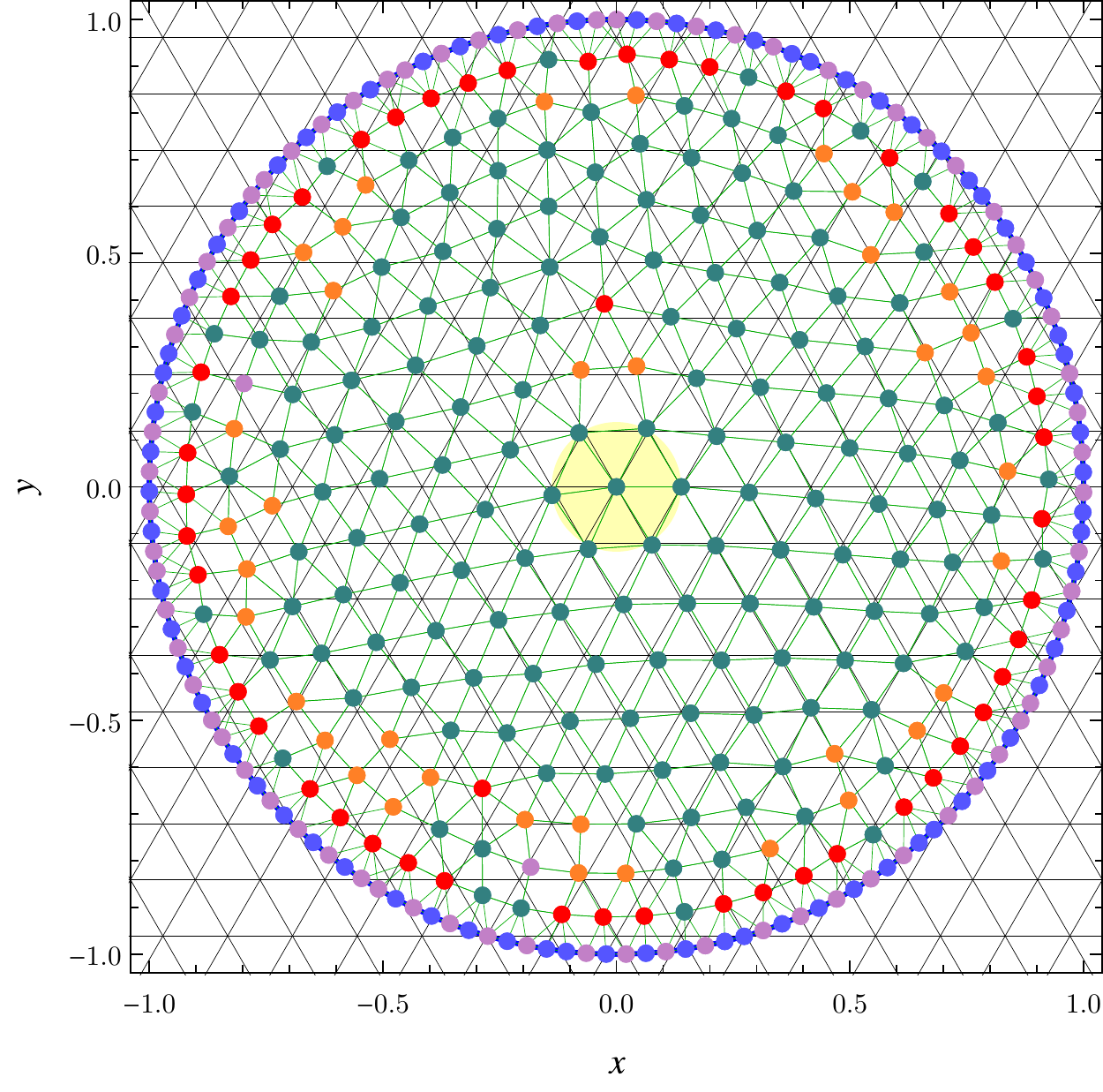} 
\caption{(color online) Numerical solution corresponding to a configuration of $395$ charges with energy
$\mathcal{E} = 110664.44$}
\label{Fig_2}
\end{center}
\end{figure}

In Fig.~\ref{Fig_2} we display the configuration with the lowest energy among those calculated (an hexagonal grid
is also plotted, to facilitate the identification of a centered hexagonal structure); note that the color of the 
vertices, representing the charges, depends on the number of nearest neighbors. Studying this figure, 
we observe the presence of defects very close to the center of the disk, and of a single, slightly deformed, 
hexagonal cell (the yellow region), centered at the origin, in sharp contrast with the numerical and theoretical observations
of Ref.~\cite{Cerkaski15}. Additionally, we have also found that similar behaviors are also observed for the 
configurations with slightly larger energy.

We summarize our main findings:
\begin{itemize}
\item the occurrence of defects, even very close to the center of the disk, may help to lower the total energy, 
while disrupting the hexagonal structure;
\item all the configurations with $N_p=147$ that we have calculated have energy lower than the theoretical 
value obtained in Ref.~\cite{Cerkaski15} and about $27 \%$ of them has energy lower than the numerical MD value of 
Ref.~\cite{Cerkaski15};
\item the use of the model of Ref.~\cite{Cerkaski15}  as a guide for the numerical MD calculation (which is
also claimed to cut the CPU times by a factor $10^3$), as suggested by the authors, is definitively 
unjustified. Using this approach, we may expect that the solutions not only will correspond to local minima
of the energy, but they will also be strongly biased (i.e. the more symmetric structures could be favored);
\item the number of configurations grows very fast with $N$, thus requiring an efficient numerical approach: 
our program allows to generate configurations with a desired number of charges on the border. We have
found that in this way the performance is drastically improved;
\end{itemize}

In light of this findings, the validity of the model of Ref.~\cite{Cerkaski15} must be questioned, particularly
for $N \gtrsim 200$; it also appear clear the inadeguacy of the numerical calculation of Ref.~\cite{Cerkaski15},
which has failed to identify a very large number of configurations with energy lower than the one reported by the 
authors (about $27 \%$ of the configurations that we have calculated have lower energy than the one of Ref.~\cite{Cerkaski15}!).
We are not sure whether  this problem has been triggered by using the output of the model as a guide for the numerical calculation
or if the program used by the authors produced a small number of configurations with $N_p=147$.

\section*{Acknowledgments}
This research  was supported by the Sistema Nacional de Investigadores (M\'exico).

\end{document}